\documentclass[onecolumn, draftcls]{IEEEtran}

\usepackage{amsbsy}
\usepackage{amsmath}
\usepackage{mathtools}
\usepackage{amsfonts} % Needed for blackboard bold, etc
\usepackage{amssymb}  % a superset of amsfonts
\usepackage{amsthm}   % for fancy theorem styles
\usepackage{amsxtra}
\usepackage{bm}
\usepackage{multirow}
\usepackage{nomencl}
\usepackage{color,colortbl}
\usepackage{dsfont}
\usepackage{enumerate}
\usepackage{fancybox}
\usepackage{fancyhdr}
\usepackage{graphics}
\usepackage{graphicx}
\usepackage{latexsym}
\usepackage{subfig}
\graphicspath{{./images/}}
\usepackage{algorithmic}

\definecolor{Dblue}{rgb}{0,0,1}
\definecolor{Dbrown}{rgb}{0.59,0.4,0}
\definecolor{Dred}{rgb}{0.64,0,0}
\definecolor{Dgreen}{rgb}{0,0.4,0}
\def\real{\mathbb R}
\def \bs {\boldsymbol}
\def \mr {\mathrm}
\def \tb {\textbf}

\def \mt {\mathrm{T}}

\newtheorem{theorem}{Theorem}[section]
\newtheorem{lemma}{Lemma}[section]
\newtheorem{definition}{Definition}[section]

\makenomenclature
\begin{document}

\title{Joint Sparse Recovery Method for Compressed Sensing with Structured Dictionary Mismatches}
\author{Zhao~Tan~\IEEEmembership{Student Member,~IEEE},
        Peng~Yang~\IEEEmembership{Student Member,~IEEE},
        and~Arye~Nehorai~\IEEEmembership{Fellow,~IEEE}
\thanks{The authors are with the Preston M. Green Department of Electrical and Systems Engineering Department, Washington University in St. Louis, St. Louis,
MO, 63130 USA. E-mail: \{tanz, yangp, nehorai\}@ese.wustl.edu.}% <-this % stops a space
\thanks{This work was supported by the AFOSR Grant FA9550-11-1-0210,
NSF Grant CCF-0963742, and ONR Grant N000141310050. }}

\maketitle

\begin{abstract}
In traditional compressed sensing theory, the dictionary matrix is given a priori, whereas in real applications this matrix suffers from random noise and fluctuations. In this paper we consider a signal model where each column in the dictionary matrix is affected by a structured noise. This formulation is common in direction-of-arrival (DOA) estimation of off-grid targets, encountered in both radar systems and array processing. We propose to use joint sparse signal recovery to solve the compressed sensing problem with structured dictionary mismatches and also give an analytical performance bound on this joint sparse recovery. We show that, under mild conditions, the reconstruction error of the original sparse signal is bounded by both the sparsity and the noise level in the measurement model. Moreover, we implement fast first-order algorithms to speed up the computing process. Numerical examples demonstrate the good performance of the proposed algorithm, and also show that the joint-sparse recovery method yields a better reconstruction result than existing methods. By implementing the joint sparse recovery method, the accuracy and efficiency of DOA estimation are improved in both passive and active sensing cases.

%\boldmath

\end{abstract}

\begin{keywords}
\noindent  compressed sensing, structured dictionary mismatch, performance bound, off-grid targets, direction-of-arrival estimation, MIMO radars, nonuniform linear arrays
\end{keywords}

\section{Introduction}
Compressed sensing is a fast growing area in the field of signal reconstruction \cite{CS1}\nocite{CS2}\nocite{CS3}-\cite{eldarbook}. It enables signal reconstruction by using a sample rate less than the normal Nyquist rate, as long as the signal of interest is sparse in a basis representation. Compressed sensing covers a wide range of applications, such as imaging \cite{CSapp1}, radar signal processing \cite{CSMIMO1}\nocite{CSMIMO2}-\cite{SandeepMIMO}, and remote sensing \cite{CSapp2}.  A typical compressed sensing problem employs the following linear model:
\begin{equation}
\bs y=\bs D\bs s+\bs w,
\end{equation}
in which $\bs D\in \real^{M\times N} (M\leq N)$ is a given dictionary matrix, $\bs y \in \real^M$ is the measurement vector, and $\bs w \in \real^M$ is the unknown noise term.  The signal of interest is $\bs s \in \real^N$, which is known to be sparse, i.e., the number of nonzero terms in $\bs s$ is far less than $N$. 

In real applications, we normally do not have perfect information about the dictionary matrix $\bs D$. The dictionary can be written as $\bs D=\bs A+\bs E$ with matrix $\bs A\in \real^{M\times N}$ known, and matrix $\bs E\in \real^{M\times N}$ unknown. In \cite{sen_mismatch}, \cite{MAB10}, the authors showed that the reconstruction error increases with the mismatch level. In this work, we consider a particular structured dictionary mismatch model with $\bs d_i=\bs a_i+\beta_i\bs b_i, 1\leq i\leq N$, where $\bs d_i$ and $\bs a_i$ are the $i$-th column of matrices $\bs D$ and $\bs A$ respectively; $\bs a_i$ and $\bs b_i$ are given for all $i$, and $\beta_i$ is unknown. Thus the signal model in our paper is
\begin{equation}\label{eq:model}
\quad \bs y=(\bs A+\bs B \bs \Delta)\bs s+\bs w,
\end{equation}
where $\bs \Delta=\mr {diag}(\bs \beta)$, $\bs \beta=[\beta_1,\beta_2,\dots,\beta_N]^\mr T$, and $\bs B=[\bs b_1,\bs b_2,\dots,\bs b_N]\in \real^{M\times N}.$

This structured mismatch was previously considered in \cite{Xie12, ETS11}. Although it is a limited mismatch model, it has many applications in areas such as spectral estimation, radar signal processing, and DOA estimation. In \cite{Xie12, zhuhao}, the alternating minimization method is proposed to solve  simultaneously for sparse signal $\bs s$ and mismatch $\bs \beta$ in \eqref{eq:model}. However, this method suffers from slow convergence and has no performance guarantee. In \cite{TYH12}, a greedy method based on matching pursuit is proposed to combine with the total least square method to deal with the structured mismatch for compressed sensing. In \cite{Xie12, ETS11}, a bounded mismatch parameter $\bs \beta$ is considered, which is common in DOA estimations for off-grid targets. The proposed frameworks were based on the first order Taylor expansion, and they enforced the sparsity of the original signal $\bs s$. They were solved using interior point methods \cite{Cvxopt}, which require solving linear systems, and the computing speed can be extremely slow when the problem's dimension grows.

In this work, we first propose to use the idea of the joint-sparse recovery \cite{Eldar09},\cite{Eldar10} to further exploit the underlying structure in compressed sensing with the structured dictionary mismatch. Joint sparsity in this paper indicates that the nonzero terms in the sparse signal come in pairs. We also give a performance guarantee when the sensing matrix $\bs A$ and the mismatch matrix $\bs B$ satisfy certain constraints. For large-dimensional problems, we implement the idea of a first-order algorithm, named fast iterative shrinkage-thresholding algorithm (FISTA) \cite{FISTA}, to solve the joint-sparse recovery with both bounded and unbounded mismatch parameter $\beta$. FISTA is a special case of a general algorithmic framework \cite{Nes13} and is more efficient in dealing with large dimensional data than the interior point methods. Some preliminary results of this work were shown in \cite{Zhaoconf}. 

We extend the developed theory and algorithms to real DOA estimation applications with both passive and active sensing. Since the number of targets in the region of interest is limited, DOA estimation  benefits from compressed sensing: both sampling energy and processing time can be greatly reduced. In order to implement compressed sensing, the region of interest needs to be discretized into a grid. The existence of off-grid targets deteriorates the performance of compressed sensing dramatically. Recent research has used compressed sensing in both active sensing application \cite{CSMIMO1}\nocite{CSMIMO2}-\cite{SandeepMIMO} and passive sensing \cite{corr,corrIM}. However, none of these works consider the situation of off-grid targets. According to the numerical example shown in this paper, by exploiting the first order derivative of sensing model associated with off-grid targets and also the joint sparsity between original signal and mismatch parameter, the accuracy of DOA estimation can be improved compared with previous methods.

The paper is organized as follows. In section \ref{sec:model} we introduce the model for compressed sensing with structured dictionary mismatches and propose to use joint sparsity to solve the reconstruction problem. We analyze the performance bound on the reconstruction error using the proposed joint sparse recovery method. In section \ref{sec:DOA} we extend the general mismatch model to the research area of DOA estimation with off-grid targets. In section \ref{sec:fista}, we give the FISTA implementation of the joint sparse recovery methods. In section \ref{sec:sensing}, we describe the mathematical model for both passive sensing and active sensing applications with off-grid targets. In section \ref{sec:numerical}, we use several numerical examples to demonstrate that the proposed method outperforms existing methods for compressed sensing with structured dictionary mismatches. Finally, in section \ref{sec:con} we conclude the paper and point out directions for future work.

We use a capital italic bold letter to represent a matrix and a lowercase italic bold letter to represent a vector. For a given matrix $\bs D$, $\bs D^\mr{T}, \bs D^*, \bs D^\mr H$ denote the transpose, conjugate transpose and conjugate without transpose of $\bs D$ respectively. For a given vector $\bs x$, $\|\bs x\|_1, \|\bs x\|_2$ are the $\ell_1$ and $\ell_2$ norms, respectively, and $\|\bs x\|_\infty$ denotes the element in $\bs x$ with the largest absolute value. Let $\|\bs x\|_0$ represent the number of nonzero components in a vector, which is referred as the $\ell_0$ norm. Let $|\bs x|$ represent a vector consisting of the absolute value of every element in $\bs x$. We use $x_i$ to represent the $i$-th element in vector $\bs x$. We use $\odot$ to denote the point-wise multiplication of two vectors with the same dimension. We use $\otimes$ to denote the Kronecker product of two matrices. In this paper, we refer a vector $\bs s$ as $K$-sparse if there are at most $K$ nonzero terms in $\bs s$.  We say a vector $\bs x \in \real^{2N}$ is $K$ joint-sparse if $\bs x=[\bs s^\mr{T}, \bs p^\mr{T}]^\mr{T}$, with $\bs s\in \real^N$ and $\bs p \in \real^N$, both being $K$ sparse with the same support set. Then we use $\|\bs x\|_{0,1}$ to denote the joint sparsity of vector $\bs x$, and we have $\|\bs x\|_{0,1}=K$ at this case.

\section{General Structured Dictionary Mismatches Model}\label{sec:model}
\subsection{Compressed Sensing with Dictionary Mismatches}
Traditional compressed sensing can be solved using the LASSO formulation \cite{lasso}, stated as
\begin{equation}
\mr{(LASSO)}\quad\min_{\bs s\in \real^n}  \frac{1}{2}\|\bs D\bs s-\bs y\|_2^2+\lambda \|\bs s\|_1.
\end{equation}
In order to recover the sparse signal $\bs s$ in the mismatch model \eqref{eq:model}, having $\bs D=\bs A+\bs B \bs \Delta$ the optimization problem is given as
\begin{equation}\label{original_opt}
\min_{\bs s\in \real^N, \bs \beta \in \real^N}  \frac{1}{2}\|(\bs A+\bs B \bs \Delta)\bs s-\bs y\|_2^2+\lambda \|\bs s\|_1, \mr{s.t.} \bs \Delta=\mr {diag}(\bs \beta).
\end{equation}
The above optimization is non-convex and generally hard to solve. Please note that when $s_i=0$ for certain $i$, then $\beta_i$ can be any value, without affecting the reconstruction. Therefore, in the rest of this paper, we focus only on instances of $\beta_i$ with nonzero $s_i$. In \cite{Xie12, zhuhao}, the authors proposed to use the alternating minimization method to solve for both $\bs s$ and $\bs \beta$ when the mismatch variable $\bs \beta$ is bounded or Gaussian distributed. Based on the idea of \cite{Xie12}, we let $\bs p=\bs \beta \odot \bs s$ and $\bs \Phi=[\bs A, \bs B]$, and then transform the original non-convex optimization into a relaxed convex one. Due to the fact that $p_i$ is zero whenever $s_i$ is zero, instead of enforcing the sparsity of $\bs s$ in \cite{Xie12, ETS11} we enforce the joint sparsity between $\bs s$ and $\bs p$. We let $\bs x=[\bs s^\mr{T}, \bs p^\mr{T}]^\mr{T} \in \real^{2N}$, and define the mixed $\ell_2/\ell_1$ norm of $\bs x$ as 
\begin{equation}
\|\bs x\|_{2,1}=\sum_{i=1}^{N}\sqrt{x_i^2+x_{N+i}^2}.
\end{equation}
Also we define 
\begin{equation}
\|\bs x\|_{\infty,1}=\max_{1\leq i\leq N}\sqrt{x_i^2+x_{N+i}^2}.
\end{equation}

If $\bs s$ is $K$-sparse, then $\bs p$ will also be $K$-sparse, with the same support set as $\bs s$. Hence the relaxed optimization enforcing joint sparsity will be referred as (JS) throughout the paper and it can be stated as
\begin{equation} \label{eq:JS}
\mr{(JS)} \quad \min_{\bs x\in \real^{2N}}  \frac{1}{2}\|\bs \Phi\bs x-\bs y\|_2^2+\lambda \|\bs x\|_{2,1}.
\end{equation}

\subsection{Performance Bound for Joint Sparse LASSO}\label{sec:bound}
In order to analyze the recovery performance of (JS), we introduce the joint restricted isometry property (J-RIP), similar to the restricted isometry property (RIP) \cite{CS1} in compressed sensing. This definition is a special case of the Block RIP introduced in \cite{Eldar09}.
\begin{definition}
{\tb{(J-RIP)}} We say that the measurement matrix $\bs \Phi \in \real^{M\times 2N}$ obeys the joint restricted isometry property with constant $\sigma_K$ if
\begin{equation}
(1-\sigma_K)\|\bs v\|_2^2 \leq \|\bs \Phi\bs v\|_2^2 \leq (1+\sigma_K)\|\bs v\|_2^2
\end{equation}
holds for all $K$ joint-sparse vectors $\bs v \in \real^{2N}$.
\end{definition}

With this definition a non-convex recovery scheme can be obtained. 
\begin{theorem}\label{thm_RIP}
Let $\bs y=\bs \Phi \bs x$, and $\bs \Phi \in \real^{M\times 2N}$, $\bs x=[\bs s^\mr T, \bs p^\mr T]^\mr T$, in which $\bs p=\bs s\odot \bs \beta \in \real^N$ and $\bs s \in \real^N$. Let $\|\bs x\|_{0,1}$ denote the joint sparsity of vector $\bs x$. Assume the matrix $\bs \Phi$ satisfies the J-RIP condition with constant $\sigma_{2K}<1$ and $\bs s$ has at most $K$ nonzero terms. By solving the following non-convex optimization problem
\begin{equation}
\min_{\bs x\in \real^{2N}}\|\bs x\|_{0,1}, \quad \mr{s.t.} \quad \bs y=\bs \Phi \bs x,
\end{equation}
we obtain the optimal solution $\hat{\bs x}$. Then $s_i=\hat{x}_i$ for all $i$, and $\beta_i=\hat{x}_{N+i}/\hat{x}_i$ when $s_i$ is nonzero.
\end{theorem}
\noindent \tb{Proof:} 
When $\bs s$ has sparsity $K$, then we know that $\|\bs x\|_{0,1}\leq K$. Then since $\hat{\bs x}$ solves the optimization problem, we have $\|\hat{\bs x}\|_{0,1} \leq \|\bs x\|_{0,1}\leq K$, and then $\|\hat{\bs x}-\bs x\|_{0,1}\leq 2K$. Since both $\hat{\bs x}$ and $\bs x$ meet the equality constraint, we have $\bs \Phi \bs x=\bs y$ and $\bs \Phi \hat{\bs x}=\bs y$, thus $\bs \Phi(\bs x-\hat {\bs x})=0$.  Using the property of J-RIP, we have
\begin{equation}
(1-\sigma_{2K})\|\bs x-\hat{\bs x}\|_2^2\leq \|\bs \Phi (\bs x-\hat{\bs x})\|_2^2=0.
\end{equation}
Hence we have $\hat{\bs x}=\bs x=[\bs s^\mr T, \bs p^\mr T]^\mr T$. Since $\bs p=\bs s \odot \bs \beta$, we than obtain $\bs s$ and $\bs \beta$ from $\hat{\bs x}$. $\square$

Since the above optimization is non-convex, the $\ell_{2,1}$ norm is used instead of the joint sparsity. Considering the noise in the signal model, the optimization takes the form
\begin{equation}\label{basis}
\min_{\bs x\in \real^{2N}}\|\bs x\|_{2,1}, \quad \mr{s.t.} \quad \|\bs y-\bs \Phi \bs x\|\leq \varepsilon.
\end{equation}
The (JS) is equivalent to the above formulation, i.e., for a given $\varepsilon$, there is a $\lambda$ that makes these two optimizations yield the same optimal point. A theoretical guarantee for \eqref{basis} is given in \cite{Eldar09}, however this result cannot be directly applied to (JS). A performance bound for (JS) can be obtained based on techniques introduced in \cite{Eldar09, Zhao13} and \cite{RIP}, and is given in the following theorem. The details of the proof is included in the Appendix.

\begin{theorem} \label{Main:thm1}
Let $\bs \Phi \in \real^{M \times 2N}$ satisfy the joint RIP with $\sigma_{2K} <0.1907$. Let the measurement $\bs y$ follow $\bs y=\bs \Phi \bs x+\bs w$, where $\bs w$ is the measurement noise in the linear system. Assume that $\lambda$ obeys $\|\bs \Phi^\mr{T}\bs w\|_{\infty,1} \leq \frac{\lambda}{2}$, and then the solution $ \hat{\bs x}$ to the optimization problem (JS) satisfies
\begin{equation}\label{mainthm}
\| \hat{\bs x}-\bs x \|_2 \leq C_0\sqrt{K} \lambda+C_1\frac{\|\bs x-(\bs x)_K\|_{2,1}}{\sqrt{K}}.
\end{equation}
Here $(\bs x)_K$ is the best K joint-sparse approximation to $\bs x$. $C_0$ and $C_1$ are constants that depend on $\sigma_{2K}$.
\end{theorem}

\noindent \tb{Remarks:}\\
\noindent \tb{1.} In \cite{Eldar09}, it was shown that random matrices satisfy the J-RIP with an overwhelming probability, and this probability is much larger than the probability of satisfying the traditional RIP under the same circumstance.\\
\noindent \tb{2.} In our case, $\bs x=[\bs s^\mr{T}, \bs p^\mr{T}]^\mr{T}$. So if $\bs s$ is $K$-sparse, since $p=\bs \beta \odot \bs s$, then $\bs x$ will be joint $K$-sparse. Thus we have $\|\bs x-(\bs x)_K\|_{2,1}=0$, and the reconstruction error depends only on the noise level, which is characterized by $\lambda$.\\
\noindent \tb{3.} In the performance bound (\ref{mainthm}), the bound is on the reconstruction error of $\bs x$, while we care more about the error bound of $\bs s$. It is easy to get 
\begin{equation}
\| \hat{\bs s}-\bs s \|_2 \leq \| \hat{\bs x}-\bs x \|_2 \leq C_0\sqrt{K} \lambda+C_1\frac{\|\bs x-(\bs x)_K\|_{2,1}}{\sqrt{K}}. 
\end{equation}
\noindent \tb{4.} In some applications, we care about $\beta_i$ only when the signal $s_i$ is nonzero. For the $i$-th element of the mismatch variable $\bs \beta$, we have 
\begin{equation}
|\hat{\beta}_i\hat{s}_i-\beta_i s_i| \leq C,
\end{equation}
where $C=C_0\sqrt{K} \lambda+C_1\frac{\|\bs x-(\bs x)_K\|_{2,1}}{\sqrt{K}}$. Using triangle inequality, we have
\begin{equation}
|\hat{s}_i||\beta_i-\hat{\beta}_i| \leq C+|\beta_i||s_i-\hat{s_i}|.
\end{equation}
When $s_i$ is nonzero, the reconstructed $\hat{s}_i$ is also highly likely to be nonzero, which is confirmed by numerical examples. In real applications, the mismatch term $\beta$ is often bounded; therefore, we can bound the reconstruction error of $\beta_i$ as
\begin{equation}
|\beta_i-\hat{\beta}_i| \leq \frac{C+|\beta_i||s_i-\hat{s_i}|}{|\hat{s}_i|}.
\end{equation}
\noindent \tb{5.} There are two ways to recover the mismatch parameter $\bs \beta$. The first way is to directly use the optimal solution from solving (JS) and let $\hat \beta_i=\hat p_i/\hat s_i$. The other way is to use the recovered $\hat {\bs s}$ from solving (JS) and plug it back in the original optimization problem (\ref{original_opt}) to solve for $\bs \beta$. 

\section{DOA Estimation with Off-grid Targets}\label{sec:DOA}
\subsection{Off-Grid Compressed Sensing}\label{offgrid}
We begin by introducing the general model encountered in DOA estimation, which is also referred as the translation-invariant model in \cite{ETS11}. The $m$th measurement in the model is described by
\begin{equation}
y_m=\sum_{k=1}^K f_k a_m(\tau_k)+w_m,
\end{equation}
where $\tau_k$ is the location of $k$th target, $w_m$ is the measurement noise and $f_k$ is the signal transmitted from $k$th target. Suppose that the region of interest spans from $\theta_1$ to $\theta_N$. Then the traditional approach is via discretizing the continuous region uniformly into a grid such as $\bs \theta=[\theta_1,\theta_2,\dots,\theta_N]$ with step size $2r$, i.e., $\theta_{i+1}-\theta_i=2r, 1\leq i\leq N-1$. Thus the signal model can be written as
\begin{equation}\label{wrong_model}
\bs y=\bs A(\bs \theta) \bs s+\bs w,
\end{equation}
where $A_{mn}(\bs \theta)=a_m(\theta_n)$, and $\bs w=[w_1,w_2,\dots, w_M]^\mr T$ is the noise term. $s_n$ is equal to $f_k$ when $\theta_n=\tau_k$ for certain $k$, otherwise $s_n$ is zero.

The model \eqref{wrong_model} is accurate only when $\tau_k \in \bs \theta$ for all $k$. When the actual parameters do not fall exactly on the discretized grid $\bs \theta$, the modeling error deteriorates the reconstruction accuracy, and the performance of compressed sensing can be highly jeopardized \cite{sen_mismatch}. Let $\bs \varphi=[\varphi_1,\varphi_2,\dots,\varphi_N]$ be the unknown grid, such that $\tau_k \in \bs \varphi$ for all $k$, and $|\varphi_n-\theta_n|\leq r$ with $1\leq n\leq N$. In this paper, we assume that two targets are at least $2r$ apart, i.e., $|\tau_i-\tau_j|>2r$ for all $1\leq i, j \leq K$. Using the first order Taylor expansion,  a more accurate signal model can be described by the unknown grid $\bs \varphi$ as
\begin{equation}\label{acc_model}
\bs y=\bs A(\bs \varphi) \bs s+\bs w \approx (\bs A+\bs B \bs \Delta)\bs s+ \bs w,
\end{equation}
where $\bs A=\bs A(\bs \theta), \bs B=[\frac{\partial \bs a(\theta_1)}{\partial \theta_1},\frac{\partial \bs a(\theta_2)}{\partial \theta_2},\dots,\frac{\partial \bs a(\theta_N)}{\partial \theta_N}], \bs \Delta=\mr{diag} {(\mr {\bs \beta})}$, and $\bs \beta=\bs \varphi-\bs \theta$. The reconstruction of the original signal $\bs s$ and grid mismatch $\bs \beta$ can be estimated by solving the (JS) optimization in \eqref{eq:JS}.

Since we know that every element in $\bs \beta$ is in the range of $[-r,r]$, one more bounded constraint can be added. By letting $\bs p=\bs \beta \odot \bs s$ and penalizing the joint sparsity between $\bs s$ and $\bs p$ we can state the non-convex bounded joint sparse method as 
\begin{eqnarray}
\mathop{\min}_{\bs s,\bs p,\bs x}  &\frac{1}{2}\|\bs A \bs s+\bs B\bs p-\bs y\|_2^2+\lambda \|\bs x\|_{2,1},\\ 
\mr{s.t.} & -r|\bs s|\leq \bs p \leq r|\bs s|,\notag \\
& \bs x=[\bs s^\mr{T},\bs p^\mr{T}]^\mr T.\notag
\end{eqnarray}

The above optimization is hard to solve. However when $\bs s$ is a positive vector, the above optimization is convex and given as 
\begin{eqnarray}\label{eq:BJS}
\mr{(BJS)} \quad \mathop{\min}_{\bs s,\bs p,\bs x}  &\frac{1}{2}\|\bs A \bs s+\bs B\bs p-\bs y\|_2^2+\lambda \|\bs x\|_{2,1},\\ 
\mr{s.t.} & -r \bs s\leq \bs p \leq r \bs s, \quad \bs s \geq 0,\notag\\
& \bs x=[\bs s^\mr{T},\bs p^\mr{T}]^\mr T.\notag
\end{eqnarray}
This formulation can be solved by standard convex optimization methods, such as interior point methods. When the dimension of the problem increases, a fast algorithm is implemented to reduce the computational burden, as we will illustrate later in this paper.

\subsection{Merging Process for Representation Ambiguity}\label{merge}
When a target is located at the midpoint of the interval $[\theta_i ,\theta_{i+1}]$ with length $2r$, then the DOA of that target can be regarded as either $\theta_i+r$ or $\theta_{i+1}-r$. This phenomenon leads to ambiguity in the reconstruction. Even in cases when the target is near the midpoint of the interval $[\theta_i, \theta_{i+1}]$, due to the measurement noise we normally have two nonzero terms of the reconstructed signal located in the interval $[\theta_i, \theta_{i+1}]$. 

To resolve this problem, we perform a linear interpolation on the two nonzero terms in the same interval and merge them into one target, since we know a priori that the two targets are at least $2r$ apart. Suppose that after solving (BJS) we have two recovered DOAs, $\varphi_a, \varphi_b \in [\theta_i,\theta_{i+1}]$. The corresponding reconstructed signal magnitudes are $s_a$ and $s_b$. After merging them, we have only one recovered DOA $\varphi$, with magnitude $s$ given as
\begin{equation}
s=s_a+s_b, \text{ and } \varphi=\theta_c+\frac{|s_a|(\varphi_a-\theta_c)+|s_b|(\varphi_b-\theta_c)}{|s_a|+|s_b|},
\end{equation}
where $\theta_c$ is the midpoint of interval $[\theta_i,\theta_{i+1}]$.

\section{Implementation with Fast First Order Algorithms}\label{sec:fista}
Using interior point methods can be time consuming for large problems. In order to speed up the computing process for (JS) and (BJS) in \eqref{eq:JS}, \eqref{eq:BJS}, we can use a first order method based on a proximal operator, namely the Fast Iterative Shrinkage-Thresholding Algorithm (FISTA) \cite{FISTA}. In this section, we first review the key concept in FISTA. The implementation of FISTA for (JS) is straightforward, while (BJS) requires more effort since it has convex constraints in the optimization problem. A smoothing function \cite{Smoothing1} is introduced to approximate $\|\bs x\|_{2,1}$ in order to implement FISTA, and continuation techniques \cite{continuation} based on the smoothing parameter are introduced to further increase the convergence speed.

\subsection{Review: FISTA and proximal operator}
To introduce the algorithm, we first review a  key concept used in FISTA, named Moreau's proximal operator, or proximal operator for short \cite{M65}. For a closed proper convex function $h:\mathbb{R}^{N} \rightarrow \mathcal{\mathbb{R}}\cup\{\infty\}$, the proximal operator of $h$ is defined by
\begin{equation}
\mr{prox}_h(\bs x)= \underset{\bs u\in \real^N}{\arg\min} \left \{h(\bs u)+\frac{1}{2}\|\bs u-\bs x\|^2_2\right \}.
\end{equation}

The proximal operator is a key step in FISTA  that solves the following composite nonsmooth problem:
\begin{equation}\label{general:model}
\quad \mathop{\min}_{\bs x\in \mathbb{R}^N} F(\bs x)=f(\bs x)+g(\bs x),
\end{equation}
where $f: \mathbb{R}^N \rightarrow \mathbb{R}$ is a smooth convex function, and it is continuously differentiable with a Lipschitz continuous gradient $L_{\nabla f}$:
\begin{equation}
\|\nabla f (\bs x)-\nabla f(\bs z) \|_2 \leq L_{\nabla f} \|\bs x-\bs z\|_2, \quad \mbox{ for all } \bs x,\bs z \in {\mathbb R}^N,
\end{equation}
\noindent  and $g: {\mathbb R}^N \rightarrow \real \cup \{\infty\}$ is continuous convex function which is possibly nonsmooth. The FISTA algorithm is given as follows. 
\\
\bigskip

%\normalsize
\begin{tabular}{l}\hline
\\
 \textbf{Fast Iterative Shrinkage-Thresholding Algorithm} \\
   \hline
   \\
    \tb{Input}: An upper bound $L\geq L_{\nabla f}$. \\
    \tb{Step 0.} Take $\bs z_1=\bs x_0, t_1=1.$\\
    \tb{Step k.} ($k\geq 1$) Compute\\
  \quad \quad \quad$\bs x_k={\rm prox}_{\frac{1}{L}g} \left ( \bs z_k-\frac{1}{L} \nabla f(\bs z_k)\right ).$\\
  \quad \quad \quad$t_{k+1}=\frac{1+\sqrt{1+4t^2_k}}{2}$.\\
  \quad \quad \quad$\bs z_{k+1}=\bs x_k+\frac{t_k-1}{t_{k+1}}(\bs x_k-\bs x_{k-1})$.\\
\\
 \hline
 \\
\end{tabular}
%\end{table}

The convergence rate of the sequence generated by FISTA is determined by the following theorem from \cite{FISTA}.
\begin{theorem}\label{fistaconverge}
Let $\{\bs x_k\}_{k \geq 0}$ be generated by FISTA, and let $\hat{\bs x}$ be an optimal solution of (\ref{general:model}). Then for any $ k \geq 1$,
\begin{equation}
 F(\bs x_k)-F(\hat{\bs x}) \leq \frac{2 L_{\nabla f} \|\bs x_0-\hat{\bs x}\|_2^2}{(k+1)^2}.
\end{equation}
\end{theorem}

\subsection{FISTA for compressed sensing with structured dictionary mismatches}

For optimization framework (JS), we know that $f(\bs x)=\frac{1}{2}\|\bs \Phi \bs x-\bs y\|_2^2$, then the Lipschitz constant is equal to $\|\bs \Phi\|_2^2$. When $g(\bs x)=\lambda \|\bs x\|_{2,1}$ and $\bs x\in \real^{2N}$, the proximal operator of $\bs x=[\bs s^\mr T,\bs p^\mr T]^\mr T$ is a group-thresholding operator defined as
\begin{align}\label{eq:proxjs}  
\mr{prox}_{\alpha g} (\{[x_i,x_{i+N}]\})=&\frac{[x_i,x_{i+N}]}{\sqrt{x_i^2+x_{i+N}^2}} \mr{max}(\sqrt{x_i^2+x_{i+N}^2}-\alpha\lambda,0), \notag \\
&1 \leq i \leq N.
\end{align}
Please note that this proximal operator yield $[0,0]$ when $x_i=x_{i+N}=0$. Hence, the algorithm using FISTA for (JS) is straightforward and summarized as follows:
\bigskip

%\normalsize
\begin{tabular}{l}\hline
\\
 \textbf{FISTA for Joint Sparse Recovery} \\
   \hline
   \\
    \tb{Input}: An upper bound $L\geq \|\bs \Phi\|_2^2$ and initial point $\bs x_0$.\\
    \tb{Step 0.} Take $\bs z_1=\bs x_0, t_1=1.$\\
    \tb{Step k.} ($k\geq 1$) Compute\\
$ \nabla f(\bs z_k)=\bs \Phi^\mr T(\bs \Phi \bs z_k-\bs y),$\\
$\bs x_k={\rm prox}_{\frac{1}{L}g} \left ( \bs z_k-\frac{1}{L} \nabla f(\bs z_k)\right ), \text{and } g(\bs u)=\lambda \|\bs u\|_{2,1},$\\
$t_{k+1}=\frac{1+\sqrt{1+4t^2_k}}{2},$\\
$\bs z_{k+1}=\bs x_k+\frac{t_k-1}{t_{k+1}}(\bs x_k-\bs x_{k-1}).$\\
\\
 \hline
 \\
\end{tabular}
%\end{table}

The FISTA implementation of (BJS) needs more work due to the positive and bounded constraints in the optimization. In order to use FISTA, we write these two convex constraints as an indicator function in the objective function. Then (BJS) is transformed into
\begin{eqnarray}
\quad \mathop{\min}_{\bs s,\bs p,\bs x}  &\frac{1}{2}\|\bs A \bs s+\bs B\bs p-\bs y\|_2^2+\lambda \|\bs x\|_{2,1}+I_\mathcal{F}(\bs s,\bs p),\\ 
\mr{s.t.} & \bs x=[\bs s^\mr{T},\bs p^\mr{T}]^\mr T \notag,
\end{eqnarray}
where $I_\mathcal{F}(\bs s,\bs p)$ is the indicator function for set $\mathcal{F}=\{\bs s\geq 0, -r\bs s\leq \bs p\leq r\bs s\}$. FISTA cannot be implemented directly since there are two nonsmooth functions, i.e., $\|\bs x\|_{2,1}$ and $I_\mathcal{F}(\bs s,\bs p)$, in the objective function.

One way to solve this issue is to approximate $h(\bs x)=\lambda \|\bs x\|_{2,1}$ by its Moreau envelope \cite{M65}, given as
\begin{equation}
h_{\mu}(\bs x) = \min_{\bs u \in \real^{2N}} \left \{ h(\bs u)+\frac{1}{2 \mu} \|\bs u-\bs x\|_2^2 \right \}.
\end{equation}
The Moreau envelope $h_\mu$ is continuously differentiable, and its gradient is equal to
\begin{equation}
\label{proxgrad} \nabla h_{\mu}(\bs x) = \frac{1}{\mu} (\bs x- {\rm prox}_{\mu h}(\bs x)),
\end{equation}
which is Lipschitz continuous with constant $1/\mu$ and can be computed using \eqref{eq:proxjs}. The smoothing approximation is more accurate with smaller $\mu$. For more details, please check \cite{Smoothing1}. 

By letting $f(\bs x)=\frac{1}{2}\|\bs \Phi \bs x-\bs y\|_2^2$ and $g(\bs x)=I_\mathcal{F}(\bs s, \bs p)$, the smoothed (BJS) can be presented as 
\begin{eqnarray}
(\mr{\mu BJS}) \quad \mathop{\min}_{\bs x}  &f(\bs x)+h_\mu(\bs x)+g(\bs x).
\end{eqnarray}
The Lipschitz constant for the gradient of $f(\bs x)+h_\mu(\bs x)$ is $\|\bs \Phi\|_2^2+\frac{1}{\mu}$. In order to implement FISTA, the proximal operator of $g(\bs x)$ is needed and can be expressed as a projection onto the set $\mathcal{F}$:
\begin{equation}
\mr{prox}_g(\bs x)=P_\mathcal{F}([\bs s^\mr T,\bs p^\mr T]^\mr T).
\end{equation}
Since the convex set $\mathcal{F}$ can be expressed as $\mathcal{F}=\bigcap_{i=1}^{N} \mathcal{F}_i$, where $\mathcal{F}_i=\{s_i\geq 0, -rs_i\leq p_i\leq rs_i\},$ the proximal operator can be computed element-wise, i.e.,
\begin{equation}
\mr{prox}_g(s_i, p_i)= P_{\mathcal{F}_i}(s_i,p_i).
\end{equation}
Here the projection from $[s_i,p_i]$ onto the two dimensional convex cone $\mathcal{F}_i$ is easy and given as follows,
\begin{equation}
    P_{\mathcal{F}_i}(s_i,p_i)=\left\{
  \begin{array}{l l}
    (s_i,p_i) & \quad -rs_i\leq p_i\leq rs_i,\\
   (0,0) & \quad \frac{s_i}{r}\leq p_i\leq -\frac{s_i}{r},\\
   c(1,r) & \quad rs_i\leq p_i, -\frac{s_i}{r}\leq p_i,\\
   c(1,-r) & \quad -rs_i\geq p_i, \frac{s_i}{r}\geq p_i,\\
  \end{array} \right.
\end{equation}
where $c=\frac{s_i+|rp_i|}{1+r^2}$. Hence the FISTA implementation for ($\mu$BJS) is given in the following.
\bigskip

%\normalsize
\begin{tabular}{l}\hline
\\
 \textbf{FISTA for $\mu$-Smoothed (BJS) Recovery} \\
   \hline
   \\
    \tb{Input}: \\ 
    \quad An upper bound $L\geq \|\bs \Phi\|_2^2+\frac{1}{\mu}$ and initial point $\bs x_0$.\\
    \tb{Step 0.} Take $\bs z_1=\bs x_0, t_1=1.$\\
    \tb{Step k.} ($k\geq 1$) Compute\\
$\quad \nabla f(\bs z_k)=\bs \Phi^\mr T(\bs \Phi \bs z_k-\bs y),$\\
$\quad \nabla h_{\mu}(\bs z_k) = \frac{1}{\mu} (\bs z_k- {\rm prox}_{\mu h}(\bs z_k)),$\\
$\quad \bs x_k=P_\mathcal{F} \left ( \bs z_k-\frac{1}{L} \nabla f(\bs z_k)-\frac{1}{L} \nabla h_\mu(\bs z_k)\right ),$\\
$\quad t_{k+1}=\frac{1+\sqrt{1+4t^2_k}}{2},$\\
$\quad \bs z_{k+1}=\bs x_k+\frac{t_k-1}{t_{k+1}}(\bs x_k-\bs x_{k-1}).$\\
\\
 \hline
 \\
\end{tabular}
%\end{table}

As we discussed earlier, smaller $\mu$ leads to better approximation accuracy. However, smaller $\mu$ incurs a larger $L$ in the algorithm, which forces the algorithm running longer to converge. The continuation technique was utilized in \cite{continuation, NESTA} to resolve this issue.  The idea of continuation is to solve ($\mu$BJS) with $\mu_1\geq \mu_2\geq \dots \geq \mu_f$ sequentially, and use the previous solution to warm start the next optimization.

\section{Passive and Active Sensing Applications}\label{sec:sensing}
\subsection{Passive Sensing: Nonuniform Linear Arrays}
The nonuniform linear array considered in this paper consists of $L$ sensors which are linearly located. We suppose the $l$th sensor is located at $d_l$. By discretizing the range of interest as $[\theta_1,\theta_2,\dots \theta_N]$, the received signal at time $t$ is given as
\begin{equation}\label{eq:ula}
\bs x(t)=\sum_{p=1}^P \alpha_p(t) \bs \phi(\theta_p)+\bs e,
\end{equation}
where $\alpha_p(t)$ is the signal transmitted with power $\sigma_p^2$ from the target at grid point $p$, with $\sigma_p$ equal to zero when there is no target at grid point $p$. $\bs \phi(\theta_p)$ is the steering vector for grid point $\theta_p$, with the $l$th element equal to $e^{\bs j(2\pi/\lambda)d_l\sin(\theta_p)}$, and $\lambda$ is the wavelength. 

We assume that all the targets are uncorrelated and that the noise is white Gaussian with noise power $\sigma_n^2$. Recent research \cite{nested,coprime} has proposed analyzing the covariance matrix of $\bs x(t)$ to increase the degrees of freedom of the original system. The covariance matrix of $\bs x$ is given as
\begin{equation}\label{eq:Rxx}
\bs R_{\bs x\bs x}=E(\bs x\bs x^*)=\sum_{p=1}^P \sigma_p^2 \bs \phi(\theta_p)\bs \phi(\theta_p)^*+\sigma_n^2 \bs I,
\end{equation}
in which $\bs I$ is an identity matrix. By vectoring the above equation, we have 
\begin{equation}
\bs y=\bs A(\bs \theta) \bs s+ \sigma^2\bs 1_n,
\end{equation}
where $\bs A(\bs \theta)=[\bs \phi(\theta_1)^\mr{H}\otimes\bs \phi(\theta_1),\dots, \bs \phi(\theta_P)^\mr{H}\otimes\bs \phi(\theta_P)],$ and $\bs s$ is a sparse signal equaling $[\sigma_1^2,\dots,\sigma_P^2]^\mr T$. We have $\bs 1_n=[\bs e_1^\mr T,\bs e_2^\mr T,\dots,\bs e_L^\mr T]^\mr T$, where $\bs e_i$ contains all zero elements except for $i$-th element, which equals one. Since $\bs s$ is a positive vector, the (BJS) formulation in \eqref{eq:BJS} can be implemented with $\bs B=[\frac{\partial (\bs \phi(\theta_1)^*\otimes \bs \phi(\theta_1))}{\partial \theta_1},\dots,\frac{\partial (\bs \phi(\theta_P)^*\otimes \bs \phi(\theta_P))}{\partial \theta_P}]$.

\subsection{Active Sensing: MIMO radar}
The MIMO radar model is based on the model introduced in \cite{CSMIMO2}. To make the paper self-contained we review the radar model in \cite{CSMIMO2} and then expand it to a general model considering off-grid targets. 

We consider a MIMO radar system with \(M_\mr T\) transmitters, \(M_\mr R\) receivers. Suppose there are \(K\) targets in the area of interest. In our case, we suppose the targets are stationary or moving very slowly compared with the sampling rate of the radar system. So the Doppler effect is neglected. The locations of transmitters and receivers are randomly generated within a disk. We consider the problem in two dimensional space using polar coordinates. The location of  the \(i\)-th transmitter is given by \([d_i^\mr t,\phi_i^\mr t]\), and the location of the \(j\)-th receiver by \([d_j^\mr r,\phi_j^\mr r]\). The region of interest is discretized into a grid. Suppose that the location of the \(p\)-th grid point is indicated by \([l_p,\theta_{p}]\). We assume that $l_\mr p \gg d_i^\mr t$ and $l_\mr p \gg d_j^\mr r$ for all $i, j$ and $p$. With this far field assumption, the distance between the $i$-th transmitter and the $p$-th grid point can be approximated as
\begin{equation}
d_{ip}^\mr t=l_p-\gamma_{ip}^\mr t,
\end{equation}
where $\gamma_{ip}^\mr t=d_i^\mr t\mr {cos}(\phi_i^\mr t-\theta_{p})$. We can also approximate the distance between the $j$-th transmitter and the $p$-th grid point as
\begin{equation}
d_{jp}^\mr r=l_p-\gamma_{jp}^\mr r,
\end{equation}
where $\gamma_{jp}^\mr r=d_j^\mr r\mr {cos}(\phi_j^\mr r-\theta_{p}).$  

Assume the transmitted signal from $i$-th transmitter is narrow band and is given as $x_i(t)e^{\bs j2\pi f_c t}, \quad i=1,...,M_\mr T$. Here  \(f_c\) indicates the transmitting frequency of the radar signal. Then the signal received by the $p$-th grid point in the scene can be written as
\begin{equation}
y_p(t)= \sum_{i=1}^{M_\mr T} x_i(t-\tau_{ip}^{\mathrm t})e^{\bs{j}2\pi f_c (t-\tau_{ip}^{\mathrm t})},\quad p=1,...,P,
\end{equation}
where $\tau_{ip}^{\mathrm t}$ represents the delay between the $i$-th transmitter and the $p$-th grid point. Therefore we can write the signal received by $j$-th receiver as
\begin{align}\label{z_con}
    z_j(t)=\sum_{p=1}^{P}\sum_{i=1}^{M_\mr T} \alpha_p  x_i(t-\tau_{ip}^{\mathrm t}-\tau_{jp}^{\mathrm r})e^{\bs{j}2\pi f_c (t-\tau_{ip}^{\mathrm t}-\tau_{jp}^{\mathrm r})},\\
\quad j=1,\dots,M_{\rm R},\notag
\end{align}
where $\tau_{jp}^{\mathrm r}$ represents the delay between the $j$-th receiver and the $p$-th grid point and $\alpha_p$ represents the refection factor if there is a target located at grid point $p$ otherwise it is zero. The term $e^{\bs j 2\pi f_c t}$ can also be known if the transmitters are synchronized and also share the same clock with each receivers. With the narrow band and far-field assumptions, we have
\begin{align}
z_j(nT)= \sum_{p=1}^{P}\sum_{i=1}^{M_\mr T} \alpha_p  x_i(nT)e^{-\bs{j}2\pi f_c (\tau_{ip}^{\mathrm t}+\tau_{jp}^{\mathrm r})},\\
\quad  j=1,\dots,M_R, \notag
\end{align}
in which $T$ is the sampling interval. The delay term in the previous equations can be calculated as $ \tau_{ip}^{\mathrm t}=  d_{ip}^{\mathrm t}/c, \tau_{jp}^{\mathrm r}=  d_{jp}^{\mathrm r}/c, $where $c$ stands for the transmission velocity of the signal.

Now we  rewrite the signal model in a sampled format which is more conventionally used for a signal processing system and write it as a matrix equation. In the following equations we neglect the sample interval $T$ for simplicity. The received signal at the \(p\)-th grid point equals
\begin{equation}\label{y}
    y_p(n)= \sum_{i=1}^{M_\mr T}x_i(n)e^{-\bs{j}\frac{2\pi f_c}{c}d_{ip}^{\mathrm t}}=e^{-\bs{j}\frac{2\pi f_c}{c}l_p}\sum_{i=1}^{M_\mr T}x_i(n)e^{\bs{j}\frac{2\pi f_c}{c}\gamma_{ip}^{\mathrm t}},
\end{equation}
where $n$ is the time index for the $n$-th sample. After expressing equation (\ref{y}) in its vector form, we have
\begin{equation} 
y_p(n)=e^{-\bs{j}\frac{2\pi f_c}{c}l_p}\bs{x}^\mr T(n)\bs{u}_p,
\end{equation}
where
\begin{equation} 
\bs{x}(n)=[x_1(n),\cdots,x_{M_\mr T}(n)]^\mr T,
\end{equation}
\begin{equation}
    \bs{u}_p=[e^{\bs{j}\frac{2\pi f_c}{c}\gamma_{1p}^{\mathrm t}},\cdots,e^{\bs{j}\frac{2\pi f_c}{c}\gamma_{M_\mr T p}^{\mathrm t}}]^\mr T.
\end{equation}

The signal received by the j-th receiver can be expressed as
\begin{equation}
    z_j(n)=\sum_{p=1}^{P}\alpha_p e^{-\bs{j}\frac{2\pi f_c}{c}l_p} e^{\bs{j}\frac{2\pi f_c}{c}\gamma_{jp}^{\mathrm r}}y_p(n),\quad j=1,\dots,M_{\mr R}.
\end{equation}

Suppose we take $L$ snapshots, and then stack all the measurements from the $j$-th receiver in one vector. We will have
\begin{equation}\label{zv}
    \bs{z}_j=\begin{pmatrix}
                   z_j(0) \\
                   \vdots \\
                   z_j(L-1) \\
                 \end{pmatrix}
                 %=\sum_{k=1}^{K}\alpha_ke^{-\bs{j}\frac{2\pi f_c}{c}dr_{jk}+\theta_{r_j}}\bs{y}_k \\
                 =\sum_{p=1}^{P}\alpha_p e^{-\bs{j}\frac{4\pi f_c}{c}l_p} e^{\bs{j}\frac{2\pi f_c}{c}\gamma_{jp}^{\mathrm r}}\bs{X} \bs{u}_p,
\end{equation}
where \(\bs{X}=[\bs{x}(0),\dots,\bs{x}(L-1)]^{\mathrm T}\).

In this linear model the sparse signal \(\bs{s}\) is given as
\begin{equation}
    s_p=\left\{
  \begin{array}{l l}
    \alpha_p e^{-\bs{j}\frac{4\pi f_c}{c}l_p} & \quad \text{if there is a target at $\theta_p$},\\
    0 & \quad \text{if there is no target}.\\
  \end{array} \right.
\end{equation}
Considering the measuring noise in the process, the received signal collected at $j$-th receiver is described as
\begin{equation}\label{Zu}
    \bs{z}_j
                 =\sum_{p=1}^{P}e^{\bs{j}\frac{2\pi f_c}{c}\gamma_{jp}^{\mathrm r}}\bs{X}\bs{u}_p s_p+
                 \bs{e}_j,
\end{equation}
in which $\bs{e}_j$ denotes the noise received by the $j$-th receiver during sampling. In our work we assume the noise is i.i.d. Gaussian.

Then we can rewrite equation (\ref{Zu}) as
\begin{equation}\label{Zuu}
    \bs{z}_j
                 =\sum_{p=1}^{P}e^{\bs{j}\frac{2\pi f_c}{c}\gamma_{jp}^{\mathrm r}}\bs{X}\bs{u}_p s_p+\bs{e}_j
                 =\bs \Psi_j\bs{s}+\bs{e}_j,
\end{equation}
in which \(\bs{s}=[s_1,\dots,s_P]^{\mathrm T}\), which indicates the locational signal, and \(\bs \Psi_j\) represents the measuring matrix for the $j$-th receiver:
\begin{equation}\label{theta}
    \bs \Psi_j=[e^{\bs{j}\frac{2\pi f_c}{c}\gamma_{j1}^{\mathrm r}}\bs{X}\bs{u}_1,\dots,e^{j\frac{2\pi f_c}{c}\gamma_{jP}^{\mathrm r}}\bs{X}\bs{u}_P].
\end{equation}

After making all these measurements, a sensing matrix is used to reduce the dimension of the problem. For the $j$-th receiver, we have a matrix $\bs \Phi_j \in \real^{M\times L}$ which is randomly generated and also satisfies the condition that $\bs \Phi_j \bs \Phi_j^\mr T=\bs I$ and $M \leq L$ The compressed data of the $j$-th receiver is given as  
\begin{equation}
\bs y_j=\bs \Phi_j\bs \Psi_j\bs s+\bs \Phi_j \bs e_j.
\end{equation}
To make the model more concise, we stack compressed data generated by all the receivers into one vector:
\begin{equation}
    \bs{y}=\begin{pmatrix}
                   \bs{y}_1 \\
                   \vdots \\
                   \bs{y}_{M_\mr R} \\
                 \end{pmatrix}
              =\bs A(\bs \theta)\bs{s}+\bs{w},
\end{equation}
where
\begin{equation}
    \bs A(\bs \theta)=\begin{pmatrix}
                  \bs \Phi_1 \bs \Psi_1 \\
                   \vdots \\
                  \bs \Phi_{M_\mr R}\bs \Psi_{M_\mr R}\\
                 \end{pmatrix},
\bs w=\begin{pmatrix}
                  \bs \Phi_1 \bs e_1 \\
                   \vdots \\
                  \bs \Phi_{M_\mr R}\bs e_{M_\mr R}\\
                 \end{pmatrix}.
\end{equation}

However, in real applications the targets' locations does not fall exactly on the grid point chosen to perform compressed sensing.  According to the idea introduced in section \ref{offgrid}, suppose the actual non-uniform grid we want to use is $\bs \varphi=[\varphi_1,\dots,\varphi_P]^\mr T$, and we need to take $\bs \beta=\bs \varphi-\bs \theta$ into consideration. Taking the derivative of the $p$-th column of matrix $\bs \Phi_j\bs \Psi_j$ with respect to $\theta_p$, we get
\begin{equation}
\bs b_{jp}=\bs{j}\frac{2\pi f_c}{c}e^{\bs{j}\frac{2\pi f_c}{c}\frac{\partial \gamma_{jp}^{\mathrm r}}{\partial \theta_p}}\bs \Phi_j\bs{X}\bs{u}_p+e^{\bs{j}\frac{2\pi f_c}{c}\gamma_{jp}^{\mathrm r}}\bs \Phi_j\bs{X}\frac{\partial \bs{u}_p}{\partial \theta_p},
\end{equation}

According to (\ref{acc_model}), the $p$-th column of matrix $\bs B$ consists of $\bs b_{jp}$ for $\forall j$, i.e. $\bs b_p=[\bs b_{1p}^\mr T,\dots,\bs b_{M_\mr R p}^\mr T]^\mr T$. We also have
\begin{equation}
\frac{\partial \bs{u}_p}{\partial \theta_p}=[\bs{j}\frac{2\pi f_c}{c}e^{\bs{j}\frac{2\pi f_c}{c}\frac{\partial \gamma_{1p}^{\mathrm t}}{\partial \theta_p}},\cdots,\bs{j}\frac{2\pi f_c}{c}e^{-\bs{j}\frac{2\pi f_c}{c}\frac{\partial \gamma_{M_\mr T p}^{\mathrm t}}{\partial \theta_p}}]^\mr T.
\end{equation}
After getting the matrix $\bs B$, (JS) optimization framework in \eqref{eq:JS} can be implemented to detect the targets' angular locations. More details will be explored in the numerical examples.

\section{Numerical Examples}\label{sec:numerical}
In this section, we present several numerical examples to show the advantages of using the joint sparse recovery method when dictionary mismatches exist in compressed sensing. In the first example, we randomly generate the data and mismatch parameters following Gaussian distributions. The measurement are obtained using model \eqref{eq:model}. FISTA-based joint sparse method and the alternating minimization method \cite{zhuhao} are considered in this case. We show that the joint sparse method provides a better reconstruction with less computational effort. In the last two examples, we compare the joint sparse method with P-BPDN \cite{Xie12} under both passive and active sensing scenarios. Please note that P-BPDN is also equivalent to the reconstruction method proposed in \cite{ETS11}.

\subsection{Randomly Generated Data}

In this numerical example we compare the FISTA-based joint-sparse method with the alternating minimization method proposed in \cite{zhuhao} when they are applied in the optimization \eqref{eq:model}. Both matrices $\bs A \in \real^{M\times N}$ and $\bs B \in \real^{M \times N}$ are randomly generated with a normal distribution with mean $0$ and standard deviation $1$. We set $N=100$. The noise term $\bs w$ is randomly generated according to a normal distribution with mean zero and standard deviation $\sigma_n=0.1$. The mismatch term $\bs \beta$ is also generated according to a normal distribution with standard deviation $\delta=1$. $\lambda$ is chosen as $10 \sigma_n\sqrt{2\log(N)}$. 

\begin{figure}[h!]
  \centering
  \includegraphics[width=0.5\textwidth]{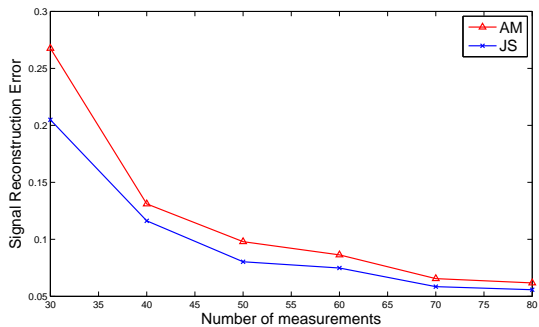}\\
  \caption{Signal reconstruction error with different number of measurements.}\label{random_M}
\end{figure}

In the first comparison, we range the number of measurements $M$ from $30$ to $80$. The sparsity of the signal $\bs s$ is 3. We use ${\|\bs s-\hat{\bs s}\|_2}/{\|\bs s\|_2}$ to denote the signal reconstruction error. We run $50$ Monte Carlo iterations at each testing point. We can see from Fig. \ref{random_M} that (JS) with FISTA performs uniformly better than the alternating minimization method. The average CPU time for alternating minimization is $15.61$s, while (JS) needs only $0.26$s.

\begin{figure}[h!]
  \centering
  \includegraphics[width=0.5\textwidth]{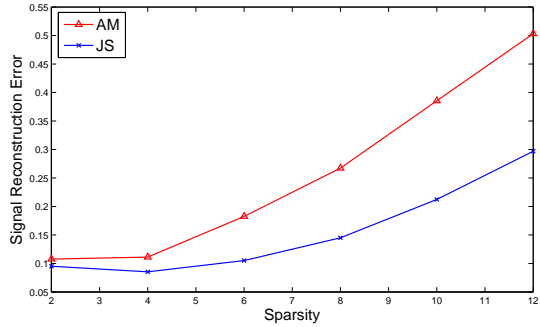}\\
  \caption{Signal reconstruction error with different sparsity level.}\label{random_K}
\end{figure}

Next, we range the sparsity level $K$ from $2$ to $12$ to compare these two methods. The number of measurements is $50$. From Fig. \ref{random_K}, we can see that (JS) has a uniformly smaller reconstruction error. The average CPU time for (JS) is $0.42$s, while the CPU time for alternating minimization is $14.34$s. 

\subsection{Nonuniform Linear Array Using Off-grid Compressed Sensing}

\begin{figure}[h!]
  \centering
  \includegraphics[width=0.5\textwidth]{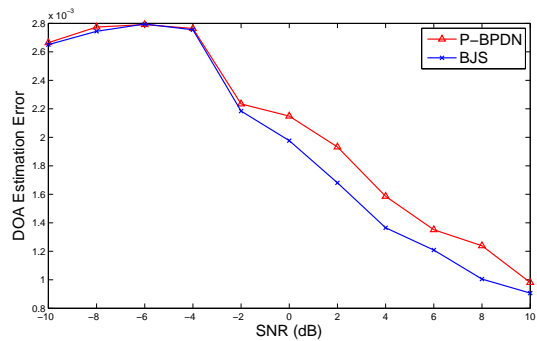}\\
  \caption{DOA estimation error with different SNR ($T=1000$).}\label{NAsnr}
\end{figure}

In this subsection, we consider a passive sensing simulation with a nonuniform linear array. The array for this part consists of two subarrays. One has sensors located at $id$ with $1\leq i\leq 5$ while the other has sensors located at $6jd$ with $1\leq j\leq 6$, and $d$ is half of the wavelength. This configuration is also called a nested array, as proposed in \cite{nested}. We compare the optimization formulation (BJS) with P-BPDN in this experiment. The power of the noise is assumed to be known; if not, an estimation of it can be easily incorporated into the (BJS) formulation. The area we are interested ranges from $\sin(\theta)=-1$ to $\sin(\theta)=1$, with a step size of $0.01$. We randomly generate $15$ targets with the same signal power. The noise at each sensor is randomly generated as white Gaussian noise with power $\sigma_n^2$. $\lambda$ in the LASSO formulation is chosen to be $\sigma_n\sqrt{2\log(N)}$ according to \cite{lambda}. However, since we use only first-order Taylor expansion to approximate the system matrix $\bs A(\bs \theta)$, the scale of the error is far larger than the additive Gaussian noise. Therefore we chose $\lambda=20\sigma_n\sqrt{2\log(N)}$ in our simulation. Here $N$ is the dimension of the signal of interest.

\begin{figure}[h!]
  \centering
  \includegraphics[width=0.5\textwidth]{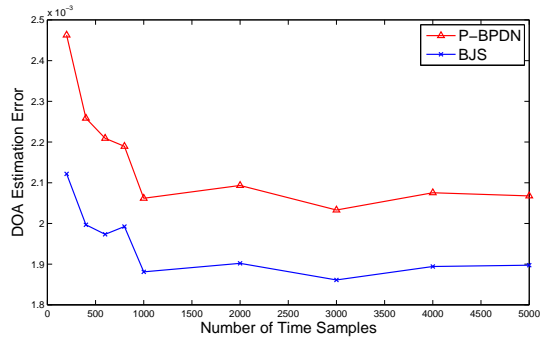}\\
  \caption{DOA estimation error with different T (SNR$=0$ dB).}\label{NAt}
\end{figure}

First we range the signal to noise ratio (SNR) from $-10$ dB to $10$ dB in Fig. \ref{NAsnr}. The number of time samples used to estimate \eqref{eq:Rxx} is $T=1000$. In Fig. \ref{NAt}, we range $T$, with the SNR fixed at $0$ dB. The DOA error is computed with respect to $\sin(\theta)$. Both figures show that (BJS) yields better DOA estimation accuracy than P-BPDN.

\begin{figure}[h!]
  \centering
  \includegraphics[width=0.5\textwidth]{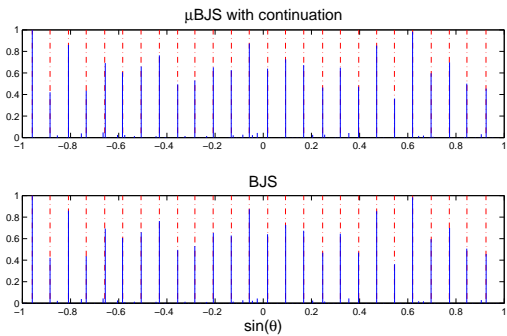}\\
  \caption{Normalized spectrum for ($\mu$BJS) with continuation, and (BJS) (T=$500$, SNR$=10$ dB).}\label{NAexample}
\end{figure}

The interior method for (BJS) works well when the dimension of the problem is small. In the next simulation, we increase the number of sensors in the linear array. The array consists of two subarrays. One has sensors located at $id$ with $1\leq i\leq 10$ while the other has sensors located at $11jd$, with $1\leq j\leq 12$. We randomly generate $26$ targets with the same signal power. We run the ($\mu$BJS) using FISTA with a continuation scheme. Let $\mu_f=10^{-8}\lambda^{-1}$. The DOA estimation results are shown in Fig. \ref{NAexample}. The running time for ($\mu$BJS) with FISTA is $4.92$s, while (BJS) with the interior point method takes $63.09$s. They both have a DOA estimation error of $5.5\times 10^{-4}$.

\subsection{MIMO Radar Using Off-grid Compressed Sensing}

In this numerical example, we compare FISTA based (JS) with P-BPDN \cite{Xie12} in a MIMO radar scenario. To fully explore the diversity of the model, we consider a MIMO system with 30 transmitters and 10 receivers whose locations are randomly generated within a disk with a radius of $5$ meters. The carrier frequency $f_c$ is $1$ GHz. Each transmitter sends out uncorrelated QPSK waveforms. The signal to noise ratio (SNR) is defined to be the ratio of the power of the transmitted waveform to the power of the additive noise in the receivers. We are interested in the area ranging from  $-40^\circ$ to $40^\circ$, with step size $1^\circ$.  We assume that two targets are at least $1^\circ$ apart.  We take $L=50$ samples for each receiver and then compress the received signal to dimension $M=10.$ Therefore we chose $\lambda=50\sigma_n\sqrt{2\log(N)}$ in our simulation.

\begin{figure}[h!]
  \centering
  \includegraphics[width=0.5\textwidth]{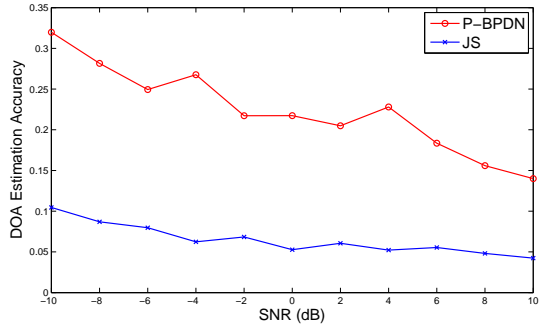}\\
  \caption{DOA estimation performance for two closely located targets with a MIMO radar system.}\label{MIMOclose}
\end{figure}

In the first simulation, we compare these two algorithms with two closely-spaced targets with SNR ranging from $-10$dB to $10$dB and show how joint sparsity benefits the reconstruction. The locations of the two targets are randomly generated from the intervals $[16.5^\circ, 17.5^\circ]$ and $[18.5^\circ, 19.5^\circ]$, with equal signal power. We run $50$ Monte Carlo iterations of every value of SNR, with the results shown in Fig. \ref{MIMOclose}. The DOA estimation error in the figure is the average DOA estimation error in degrees. We can see that the method proposed in this paper has consistent better reconstruction performance than P-BPDN for location estimation.

\begin{figure}[h!]
  \centering
  \includegraphics[width=0.5\textwidth]{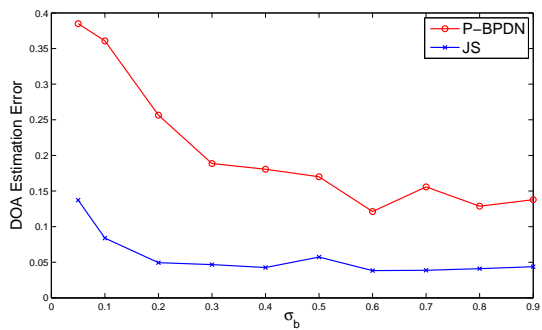}\\
  \caption{DOA estimation error with changing $\sigma_b$ $(\sigma_a=1)$.}\label{DR}
\end{figure}

In the next simulation, we compare (JS) using FISTA with P-BPDN when the dynamic range changes between these two targets. Suppose the first target is randomly generated with signal power $\sigma_a^2=1$, and the second target has a signal power $\sigma_b^2$. SNR is chosen to be $10$ dB in this case. From Fig. \ref{DR} we can see that (JS) performs better with respect to changing dynamic range.

\section{Conclusion} \label{sec:con}
In this paper, we proposed a method to overcome structured dictionary mismatches in compressed sensing. We utilized the joint sparse recovery model and also gave a performance bound on the joint sparse reconstruction. For off-grid compressed sensing, a bounded joint sparse recovery method was utilized.  Fast algorithms based on FISTA were given to solve these joint sparse recovery formulations. One important application of this framework is called off-grid compressed sensing for DOA estimation. Both passive and active sensing applications were used to demonstrate the effectiveness of the proposed algorithms. Numerical examples were conducted to compare the performance of the joint sparse method and other existing methods. We demonstrated that by exploiting the joint sparse property, we can get more satisfactory results when structured dictionary mismatches exist. In future work, we will apply our method to other applications, analyze the DOA estimation accuracy with respect to the condition of the sensing matrix, and develop a more theoretical way to choose $\lambda$ for off-grid compressed sensing.
%------------------------------------------------------------------------
%   References
% ------------------------------------------------------------------------

\appendix

Before beginning the proof of the main theorem, we give several useful lemmas which will be used in the main proof. The first lemma is based on the J-RIP property of the matrix $\bs \Phi$.
\begin{lemma}\label{DRIP}
If the matrix $\bs \Phi$ satisfies J-RIP with parameter $\sigma_{2K}$, then for all $\bs u, \bs v \in \real^{2N}$, which are both $K$ joint-sparse with non-overlapping support sets, we have
\begin{equation*}
 \langle \bs \Phi\bs u,\bs \Phi\bs v\rangle \geq -\sigma_{2K} \|\bs u\|_2\|\bs v\|_2.
\end{equation*}
\end{lemma}

\noindent \tb{Proof:}
We first consider the case when $\|\bs u\|_2=1$ and $\|\bs v\|_2=1$. According to the definition of J-RIP, we have the following inequality:
\begin{align*}
\langle \bs \Phi\bs u,\bs \Phi\bs v\rangle=&\frac{1}{4}\{\|\bs \Phi\bs u+\bs \Phi\bs v\|^2_2-\|\bs \Phi\bs u-\bs \Phi \bs v\|^2_2\}\notag \\
\geq& \frac{1}{4}\{(1-\sigma_{2K})\|\bs u+\bs v\|^2_2-(1+\sigma_{2K})\|\bs u-\bs v\|^2_2\}\notag \\
\geq& -\sigma_{2K}+\bs u^{\mr T}\bs v\notag\\
 =& -\sigma_{2K}.
\end{align*}
The last equality utilizes the fact that $\bs u$ and $\bs v$ has non-overlapping support sets. Now it is easy to extend this equation to get the result in lemma \ref{DRIP}. 

We use $\hat{\bs x}$ to represent the optimal solution of (JS) and denote $\bs x$ as the original signal with $\bs y=\bs \Phi \bs x+\bs w$; we also use $\bs h$ to represent the reconstruction error $\hat{\bs x}-\bs x$.  Now let $\mathcal{T}$ denote the index of coefficients with $k$ largest joint-magnitudes of vector $\bs x$, i.e., the indices $i$ and $N+i$ for $(1\leq i\leq N)$ with $k$ largest $\sqrt{x_i^2+x_{N+i}^2}$. $\mathcal{T}^c$ denotes the complement of $\mathcal{T}$.  Let $\bs x_{\mathcal{T}}$ be a vector that maintains the same coefficients as $\bs x$ with support set $\mathcal{T}$, while setting other indices as zeros. Let $\mathcal{T}_0=\mathcal{T}$, and we decompose $\mathcal{T}_0^c$ into sets of size $K$. Let $\mathcal{T}_1$ denote the locations of the $K$ largest joint-magnitudes in $\bs h_{\mathcal{T}^c}$, $\mathcal{T}_2$ denote the next $K$ largest joint-magnitudes in $\bs  h_\mathcal{T}^c$ and so on. We also have $\mathcal{T}_{01}=\mathcal{T}_0\cup \mathcal{T}_1$. The next lemma relates the $\ell_2$ norm of the tail to the $\ell_2/\ell_1$ norm of the tail.

\begin{lemma}\label{tail} \tb{(Bounding the tail)}
For the reconstruction error $\bs h$ from solving (JS) and disjointed sets $\mathcal{T}_0,\mathcal{T}_1,\dots$ defined earlier, we have
$$\sum_{j\geq 2}\|\bs h_{\mathcal{T}_{j}}\|_2 \leq K^{-\frac{1}{2}}\|\bs h_{\mathcal{T}^c}\|_{2,1}.$$
\end{lemma}

\noindent \tb{Proof :}
First we can write the following inequality as
$$\|\bs h_{\mathcal{T}_j}\|_2\leq K^{\frac{1}{2}}\|\bs h_{\mathcal{T}_j}\|_{\infty,1}\leq K^{-\frac{1}{2}}\|\bs h_{\mathcal{T}_{j-1}}\|_{2,1}.$$
The above equation utilizes the definition of $\|\bs x\|_{\infty,1}$ and also the fact that every joint magnitude in set $\mathcal{T}_j$ is no larger than every joint magnitude in set $\mathcal{T}_{j-1}$. By summing up over $j$, we obtain
\begin{equation}\label{boundtail}
\sum_{j\geq 2}\|\bs h_{\mathcal{T}_{j}}\|_2 \leq K^{-\frac{1}{2}} \sum_{j\geq 1}\|\bs h_{\mathcal{T}_{j}}\|_{2,1} = K^{-\frac{1}{2}}\|\bs h_{\mathcal{T}^c}\|_{2,1}.
\end{equation}
 
The lemmas below are derived from the optimality of $\hat {\bs x}$, and they show that the reconstruction error $\bs h$ and $\|\bs h_{\mathcal{T}^c}\|_{2,1}$ is bounded when $\hat{\bs x}$ solves the (JS).
\begin{lemma}\label{optimalcon}\tb{(Optimality condition 1)} 
Assuming that $\lambda$ obeys $\|\bs \Phi^\mr{T}\bs w\|_{\infty,1} \leq \frac{\lambda}{2}$, the reconstruction error $\bs h$ of (JS) satisfies the following inequality
$$\|\bs \Phi^\mt \bs \Phi \bs h\|_{\infty,1}\leq \frac{3}{2}\lambda,$$
\end{lemma}

\noindent \tb{Proof}: The optimality condition for (JS) requires that the gradients vanish to zero, and it can be stated as
\begin{equation}\label{sub1}
\bs \Phi^\mt(\bs \Phi\bs \hat{\bs x}-\bs y)+\lambda \bs v=0,
\end{equation}
where $\bs v$ is the gradient of function $\|\bs v\|_{2,1}$. It is easy to verify that $\|\bs v\|_{\infty,1}\leq 1$, so we get
\begin{align}
\|\bs \Phi^\mt(\bs \Phi \hat{\bs x}-\bs y)\|_{\infty,1}= \lambda \| \bs v\|_{\infty,1} \leq \lambda.
\end{align}
From the assumption $\|\bs \Phi^\mr T\bs w\|_{\infty,1} \leq \frac{\lambda}{2}$,
\begin{equation*}
\|\bs \Phi^\mt \bs \Phi \bs h\|_{\infty,1} \leq \|\bs \Phi^\mt (\bs \Phi \bs x-\bs y)\|_{\infty,1}+\|\bs \Phi^\mt(\bs \Phi\hat{\bs x}-\bs y)\|_{\infty,1}\leq \frac{3}{2}\lambda.
\end{equation*}

\begin{lemma}\label{cone} \tb{(Optimality condition 2)} 
For the reconstruction of (JS), we have following inequality:
\begin{equation*}
\|\bs h_{\mathcal{T}^c}\|_{2,1} \leq 3\|\bs h_{\mathcal{T}}\|_{2,1}+4\|\bs x_{\mathcal{T}^c}\|_{2,1}.
\end{equation*}
\end{lemma}

\noindent \tb{Proof}: Now, since $\hat{\bs x}$ solves the optimization problem (JS), we have
\begin{align*}
&\frac{1}{2}\|\bs \Phi \hat{\bs x}-\bs y\|^2_2+\lambda \|\hat{\bs x}\|_{2,1} \leq \frac{1}{2}\|\bs \Phi\bs x-\bs y\|_2^2+\lambda \|\bs  x\|_{2,1}.
\end{align*}
Since $\bs y=\bs \Phi \bs x+\bs w$, and by letting $\bs h$ denote $\hat{\bs x}-\bs x$, we have
\begin{align*}
\frac{1}{2}\|\bs \Phi\bs h-\bs w\|_2^2+\lambda \|\hat{\bs x}\|_{2,1} \leq \frac{1}{2}\|\bs w\|_2^2+\lambda \|\bs x\|_{2,1}.
\end{align*}
Expanding the first term on the left side and rearranging the terms in the above equation, we get
\begin{align*}
\frac{1}{2}\|\bs \Phi\bs h\|_2^2+\lambda\|\hat{\bs x}\|_{2,1} \leq &\langle \bs \Phi \bs h,\bs w \rangle+\lambda \|\bs x\|_{2,1}
\\ \leq& \|\bs \Phi^\mt \bs w\|_{\infty,1}\|\bs h\|_{2,1}+\lambda \|\bs x\|_{2,1}.
\end{align*}
The second inequality follows from the fact that $\langle \bs x, \bs y \rangle =\sum_{i=1}^N (x_iy_i+x_{N+i}y_{N+i})\leq \sum_{i=1}^N\sqrt{x_i^2+x_{N+i}^2}\sqrt{y_{i}^2+y_{N+i}^2} \leq \|\bs x\|_{2,1}\|\bs y\|_{\infty,1}$. With the assumption that $\|\bs \Phi^\mt \bs w\|_{\infty,1} \leq \frac{\lambda}{2}$, we get
\begin{align*}
\frac{1}{2}\|\bs \Phi\bs h\|_2^2+\lambda\|\hat{\bs x}\|_{2,1} \leq  \frac{\lambda}{2}\|\bs h\|_{2,1}+\lambda \|\bs x\|_{2,1}.
\end{align*}
Therefore we have
\begin{align*}
\lambda\|\hat{\bs x}\|_{2,1}\leq& \frac{1}{2}\|\bs \Phi \bs h\|_2^2+\lambda \|\bs \hat{\bs x}\|_{2,1} \leq  \frac{\lambda}{2}\|\bs h\|_{2,1}+\lambda \|\bs x\|_{2,1}.
\end{align*}
Since we have $\bs h=\hat{\bs x}-\bs x$, we also have
\begin{equation*}
\|\bs h+ \bs x\|_{2,1} \leq \frac{1}{2}\|\bs h\|_{2,1}+\|\bs x\|_{2,1}.
\end{equation*}
Using the above equation, we can show that
\begin{align*}
&\|\bs h_{\mathcal{T}}+ \bs x_{\mathcal{T}}\|_{2,1}+\|\bs h_{\mathcal{T}^c}+ \bs x_{\mathcal{T}^c}\|_{2,1} 
\\ \leq&\frac{1}{2}\|\bs h_{\mathcal{T}}\|_{2,1}+\frac{1}{2}\|\bs h_{\mathcal{T}^c}\|_{2,1}+\|\bs x_{\mathcal{T}}\|_{2,1}+\|\bs x_{\mathcal{T}^c}\|_{2,1}.
\end{align*}
Applying triangle inequality on the left hand side of above inequality, we have
\begin{align*}
&-\|\bs h_{\mathcal{T}}\|_{2,1}+ \|\bs x_{\mathcal{T}}\|_{2,1}+\|\bs h_{\mathcal{T}^c}\|_{2,1}-\| \bs x_{\mathcal{T}^c}\|_{2,1}
\\ \leq& \frac{1}{2}\|\bs h_{\mathcal{T}}\|_{2,1}+\frac{1}{2}\|\bs h_{\mathcal{T}^c}\|_{2,1}+\|\bs x_{\mathcal{T}}\|_{2,1}+\|\bs x_{\mathcal{T}^c}\|_{2,1}.
\end{align*}
After rearranging the terms, we have the following cone constraint:
\begin{equation}\label{Dhrelation}
\|\bs h_{\mathcal{T}^c}\|_{2,1} \leq 3\|\bs h_{\mathcal{T}}\|_{2,1}+4\|\bs x_{\mathcal{T}^c}\|_{2,1}.
\end{equation}

With the above lemmas, we can prove theorem \ref{Main:thm1} as follows.

\noindent \tb{Main Proof: } The proof follows some techniques in \cite{Eldar09}, \cite{Zhao13} and \cite{RIP}. The challenge lies in two aspects. First, instead of dealing with sparsity, we have to use the property of joint-sparsity for the derivation. Second, unlike the constrained optimization considered in \cite{Eldar09}, in this work we are trying to derive the performance bound for an unconstrained optimization. The proof is derived in two steps. First, we show that $\bs h$ inside the set $\mathcal{T}_{01}$ is bounded by the terms of $h$ outside the set $\mathcal{T}$. Then we show that $\bs h_{\mathcal{T}^c}$ is essentially small. First we have
\begin{align*}
\langle \bs \Phi\bs h,\bs \Phi\bs h_{\mathcal{T}_{01}}\rangle=& \langle \bs \Phi\bs h_{\mathcal{T}_{01}},\bs \Phi\bs h_{\mathcal{T}_{01}}\rangle +\sum_{j\geq 2}\langle \bs \Phi \bs h_{\mathcal {T}_{j}},\bs \Phi\bs h_{\mathcal{T}_{01}}\rangle
\\ \geq& (1-\sigma_{2K}) \|\bs h_{\mathcal {T}_{01}}\|^2_2+\sum_{j\geq 2}\langle \bs \Phi \bs h_{\mathcal {T}_{j}},\bs \Phi \bs h_{\mathcal{T}_{0}}\rangle
\\ &+\sum_{j\geq 2} \langle \bs \Phi \bs h_{\mathcal {T}_{j}},\bs \Phi \bs h_{\mathcal{T}_{1}}\rangle
\\ \geq&(1-\sigma_{2K}) \| \bs h_{\mathcal {T}_{01}}\|^2_2-\sigma_{2K}\|\bs h_{\mathcal{T}_{0}}\|_2\sum_{j\geq 2}\|\bs h_{\mathcal{T}_{j}}\|_2\notag
\\&-\sigma_{2K}\|\bs h_{\mathcal{T}_{1}}\|_2\sum_{j\geq 2}\|\bs h_{\mathcal{T}_{j}}\|_2 \notag
\\=&(1-\sigma_{2K}) \|\bs h_{\mathcal {T}_{01}}\|^2_2 \notag \\
&-\sigma_{2K}(\|\bs h_{\mathcal{T}_{0}}\|_2+\|\bs h_{\mathcal{T}_{1}}\|_2)\sum_{j\geq 2}\|\bs h_{\mathcal{T}_{j}}\|_2
\\ \geq&(1-\sigma_{2K}) \|\bs h_{\mathcal {T}_{01}}\|^2_2-\sqrt{2}\sigma_{2K}\|\bs h_{\mathcal{T}_{01}}\|_2\sum_{j\geq 2}\|\bs h_{\mathcal{T}_{j}}\|_2.
\end{align*}
The first inequality follows the J-RIP of matrix $\bs \Phi$. The second inequality uses the result from Lemma \ref{DRIP}. The third one is deduced from the fact that $\|\bs h_{\mathcal{T}_{0}}\|_2+\|\bs h_{\mathcal{T}_{1}}\|_2 \leq \sqrt{2}\|\bs h_{\mathcal{T}_{01}}\|_2 $ when the set $\mathcal{T}_0$ and the set $\mathcal{T}_1$ are disjoint.
With the result from Lemma \ref{tail}, we have our final inequality as
\begin{equation}\label{ADADDh1}
\langle \bs \Phi\bs h,\bs \Phi \bs h_{\mathcal{T}_{01}}\rangle \geq(1-\sigma_{2K}) \|\bs h_{\mathcal {T}_{01}}\|^2_2 -\sqrt{2}K^{-\frac{1}{2}}\sigma_{2K}\|\bs h_{\mathcal{T}_{01}}\|_2 \|\bs h_{\mathcal{T}^c}\|_{2,1}.
\end{equation}

From the inequality $\langle \bs x, \bs y \rangle =\sum_{i=1}^N (x_iy_i+x_{N+i}y_{N+i})\leq \sum_{i=1}^N\sqrt{x_i^2+x_{N+i}^2}\sqrt{y_{i}^2+y_{N+i}^2} \leq \|\bs x\|_{2,1}\|\bs y\|_{\infty,1}$, we get
\begin{align}\label{ADADDh2}
\langle \bs \Phi\bs h,\bs \Phi \bs h_{\mathcal{T}_{01}}\rangle =&\langle \bs \Phi^\mt \bs \Phi\bs h, \bs h_{\mathcal{T}_{01}}\rangle \leq\|\bs \Phi^\mt \bs \Phi\bs h\|_{\infty,1}\|\bs h_{\mathcal{T}_{01}}\|_{2,1}\notag \\
\leq&\sqrt{2K}\|\bs \Phi^\mt \bs \Phi\bs h\|_{\infty,1}\|\bs h_{\mathcal{T}_{01}}\|_2 \leq \sqrt{K}c_0\lambda\|\bs h_{\mathcal{T}_{01}}\|_2,
\end{align}
where $c_0=\frac{3\sqrt{2}}{2}$. The second inequality uses the fact that $\|\bs h_{\mathcal{T}_{01}}\|_{2,1}\leq \sqrt{2K}\|\bs h_{\mathcal{T}_{01}}\|_2$, which is derived by using Cauchy-Schwarz inequality. The last inequality follows the result of Lemma \ref{optimalcon}. Combining equations (\ref{ADADDh1}) and (\ref{ADADDh2}), we get
\begin{equation}\label{Dh2}
\|\bs h_{\mathcal{T}_{01}}\|_2 \leq \frac{\sqrt{K}\lambda c_0+\sqrt{2}K^{-\frac{1}{2}}\sigma_{2K}\|\bs h_{\mathcal{T}^c}\|_{2,1}}{1-\sigma_{2K}}.
\end{equation}
Hence, combining the Cauchy-Schwarz inequality with the result from last inequality leads to
\begin{align}\label{Dh3}
\|\bs h_{\mathcal{T}}\|_{2,1} \leq & \sqrt{K} \|\bs h_{\mathcal{T}}\|_2 \leq \sqrt{K} \|\bs h_{\mathcal{T}_{01}}\|_2 \notag \\
\leq & \frac{\lambda Kc_0+\sqrt{2}\sigma_{2K}\|\bs h_{\mathcal{T}^c}\|_{2,1}}{1-\sigma_{2K}}.
\end{align}

Next, we prove that $\bs h_{\mathcal{T}^c}$ is relatively small. Combining the inequalities from Lemma \ref{cone} and (\ref{Dh3}), we have
\begin{equation*}
\|\bs h_{\mathcal{T}^c}\|_{2,1} \leq \frac{3\lambda Kc_0+3\sqrt{2}\sigma_{2K}\|\bs h_{\mathcal{T}^c}\|_{2,1}}{1-\sigma_{2K}}+4\|\bs x_{\mathcal{T}^c}\|_{2,1}.
\end{equation*}
From the assumption $\sigma_{2K}<0.1907$, we have $1-(1+3\sqrt{2})\sigma_{2K}>0$. Then by rearranging the terms, the above inequality becomes
\begin{equation}\label{Dh4}
\|\bs h_{\mathcal{T}^c}\|_{2,1} \leq \frac{3\lambda Kc_0+4(1-\sigma_{2K})\|\bs x_{\mathcal{T}^c}\|_{2,1}}{1-(1+3\sqrt{2})\sigma_{2K}}.
\end{equation}
Now we can bound the reconstruction error $\bs h$. Using the results from Lemma \ref{tail} and equations (\ref{Dh2}) and (\ref{Dh4}), we derive
\begin{align}
\|\bs h\|_2 \leq& \|\bs h_{\mathcal{T}_{01}}\|_2+\sum_{j\geq 2}\|\bs h_{\mathcal{T}_{j}}\|_2 \notag \\
\leq&  \frac{\sqrt{K}\lambda c_0+\sqrt{2}K^{-\frac{1}{2}}\sigma_{2K}\|\bs h_{\mathcal{T}^c}\|_{2,1}}{1-\sigma_{2K}}+K^{-\frac{1}{2}}\|\bs h_{\mathcal{T}^c}\|_{2,1} \notag\\
=&\frac{c_0\lambda\sqrt{K}}{1-\sigma_{2K}}+\frac{((\sqrt{2}-1)\sigma_{2K}+1)K^{-\frac{1}{2}}\|\bs h_{\mathcal{T}^c}\|_{2,1}}{1-\sigma_{2K}}\notag \\
\leq& C_0\sqrt{K} \lambda+C_1\frac{\|\bs x-(\bs x)_K\|_{2,1}}{\sqrt{K}}.
\end{align}
The first inequality uses the triangle inequality. For the second inequality we use Lemma \ref{tail}. The constants are given as
\begin{align*}
C_0=\frac{6\sqrt{2}}{1-(1+3\sqrt{2})\sigma_{2K}}, \quad C_1=\frac{4((\sqrt{2}-1)\sigma_{2K}+1)}{1-(1+3\sqrt{2})\sigma_{2K}}.
\end{align*}

%------------------------------------------------------------------------
%   References
% ------------------------------------------------------------------------

\bibliographystyle{IEEEtran}
\bibliography{IEEEabrv,offgrid}

\printnomenclature
\bibliographystyle{plain}

\end{document}